\documentclass[11pt,a4paper]{article}
\input amssymb.sty
\begin{document}
\newcommand{\Si}{\Sigma}
\newcommand{\tr}{{\rm tr}}
\newcommand{\ad}{{\rm ad}}
\newcommand{\Ad}{{\rm Ad}}
\newcommand{\ti}[1]{\tilde{#1}}
\newcommand{\om}{\omega}
\newcommand{\Om}{\Omega}
\newcommand{\de}{\delta}
\newcommand{\al}{\alpha}
\newcommand{\te}{\theta}
\newcommand{\vth}{\vartheta}
\newcommand{\be}{\beta}
\newcommand{\la}{\lambda}
\newcommand{\La}{\Lambda}
\newcommand{\D}{\Delta}
\newcommand{\ve}{\varepsilon}
\newcommand{\ep}{\epsilon}
\newcommand{\vf}{\varphi}
\newcommand{\vfh}{\varphi^\hbar}
\newcommand{\vfe}{\varphi^\eta}
\newcommand{\fh}{\phi^\hbar}
\newcommand{\fe}{\phi^\eta}
\newcommand{\G}{\Gamma}
\newcommand{\ka}{\kappa}
\newcommand{\ip}{\hat{\upsilon}}
\newcommand{\Ip}{\hat{\Upsilon}}
\newcommand{\ga}{\gamma}
\newcommand{\ze}{\zeta}
\newcommand{\si}{\sigma}

\def\clA{\mathcal{A}}
\def\clG{\mathcal{G}}
\def\clR{\mathcal{R}}
\def\clU{\mathcal{U}}
\def\clO{\mathcal{O}}
\def\clL{\mathcal{L}}
\def\clZ{\mathcal{Z}}

\def\bfa{{\bf a}}
\def\bfb{{\bf b}}
\def\bfc{{\bf c}}
\def\bfd{{\bf d}}
\def\bfe{{\bf e}}
\def\bff{{\bf f}}
\def\bfm{{\bf m}}
\def\bfn{{\bf n}}
\def\bfp{{\bf p}}
\def\bfu{{\bf u}}
\def\bfv{{\bf v}}
\def\bft{{\bf t}}
\def\bfx{{\bf x}}
\def\bfg{{\bf g}}
\def\bfC{{\bf C}}
\def\bfS{{\bf S}}
\def\bfJ{{\bf J}}
\def\bfI{{\bf I}}
\def\bfP{{\bf P}}
\def\bfr{{\bf r}}
\def\bfU{{\bf U}}

\def\bfal{\breve{\al}}
\def\bfbe{\breve{\be}}
\def\bfga{\breve{\ga}}
\def\bfnu{\breve{\nu}}
\def\bfsi{\breve{\sigma}}

\def\hS{{\hat{S}}}

\newcommand{\li}{\lim_{n\rightarrow \infty}}
\def\mapright#1{\smash{
\mathop{\longrightarrow}\limits^{#1}}}

\newcommand{\mat}[4]{\left(\begin{array}{cc}{#1}&{#2}\\{#3}&{#4}
\end{array}\right)}
\newcommand{\thmat}[9]{\left(
\begin{array}{ccc}{#1}&{#2}&{#3}\\{#4}&{#5}&{#6}\\
{#7}&{#8}&{#9}
\end{array}\right)}

\newcommand{\thch}[4]{\theta\left[
\begin{array}{c}{#1}\\{#2}
\end{array}
\right]({#3};{#4})}

\newcommand{\beq}[1]{\begin{equation}\label{#1}}
\newcommand{\eq}{\end{equation}}
\newcommand{\beqn}[1]{\begin{eqnarray}\label{#1}}
\newcommand{\eqn}{\end{eqnarray}}
\newcommand{\p}{\partial}
\newcommand{\di}{{\rm diag}}
\newcommand{\oh}{\frac{1}{2}}
\newcommand{\su}{{\bf su_2}}
\newcommand{\uo}{{\bf u_1}}
\newcommand{\SL}{{\rm SL}(2,{\mathbb C})}
\newcommand{\GLN}{{\rm GL}(N,{\mathbb C})}
\newcommand{\PGLN}{{\rm PGL}(N,{\mathbb C})}

\def\sln{{\rm sl}(N, {\mathbb C})}
\def\sl2{{\rm sl}(2, {\mathbb C})}
\def\SLN{{\rm SL}(N, {\mathbb C})}
\def\SLT{{\rm SL}(2, {\mathbb C})}
\newcommand{\gln}{{\rm gl}(N, {\mathbb C})}
\newcommand{\PSL}{{\rm PSL}_2( {\mathbb Z})}
\def\f1#1{\frac{1}{#1}}
\def\lb{\lfloor}
\def\rb{\rfloor}
\def\sn{{\rm sn}}
\def\cn{{\rm cn}}
\def\dn{{\rm dn}}
\newcommand{\rar}{\rightarrow}
\newcommand{\upar}{\uparrow}
\newcommand{\sm}{\setminus}
\newcommand{\ms}{\mapsto}
\newcommand{\bp}{\bar{\partial}}
\newcommand{\bz}{\bar{z}}
\newcommand{\bw}{\bar{w}}
\newcommand{\bA}{\bar{A}}
\newcommand{\bL}{\bar{L}}
\newcommand{\btau}{\bar{\tau}}

\newcommand{\Sh}{\hat{S}}
\newcommand{\vtb}{\theta_{2}}
\newcommand{\vtc}{\theta_{3}}
\newcommand{\vtd}{\theta_{4}}

\def\mC{{\mathbb C}}
\def\mZ{{\mathbb Z}}
\def\mR{{\mathbb R}}
\def\mN{{\mathbb N}}
\def\mP{{\mathbb P}}

\def\frak{\mathfrak}
\def\gg{{\frak g}}
\def\gJ{{\frak J}}
\def\gS{{\frak S}}
\def\gL{{\frak L}}
\def\gG{{\frak G}}
\def\gk{{\frak k}}
\def\gK{{\frak K}}
\def\gl{{\frak l}}
\def\gh{{\frak h}}
\def\gH{{\frak H}}

\newcommand{\ran}{\rangle}
\newcommand{\lan}{\langle}
\def\f1#1{\frac{1}{#1}}
\def\lb{\lfloor}
\def\rb{\rfloor}
\newcommand{\slim}[2]{\sum\limits_{#1}^{#2}}

\newcommand{\sect}[1]{\setcounter{equation}{0}\section{#1}}
\renewcommand{\theequation}{\thesection.\arabic{equation}}
\newtheorem{predl}{Proposition}[section]
\newtheorem{defi}{Definition}[section]
\newtheorem{rem}{Remark}[section]
\newtheorem{cor}{Corollary}[section]
\newtheorem{lem}{Lemma}[section]
\newtheorem{theor}{Theorem}[section]


\vspace{10mm}
\begin{center}
{\Large{\bf Quadratic algebras related to elliptic curves }
}\\
\vspace{5mm}

Yu.Chernyakov,$\dag$ A.M.Levin,$\diamondsuit\ddag$ M.Olshanetsky,$\dag$\\
 A.Zotov$\dag\ddag$\\

\vspace{3mm}

{\it
$\dag$ - Institute of Theoretical and Experimental Physics, Moscow,\\
$\diamondsuit$ - Institute of Oceanology, Moscow,\\
$\ddag$ - Max Planck Institute of Mathematics, Bonn\\
}

\vspace{5mm}
\end{center}

\begin{abstract}
We construct quadratic finite-dimensional Poisson algebras and
their quantum versions related to rank N and degree one vector
bundles over elliptic curves with n marked points. The algebras
are parameterized by the moduli of curves. For N=2 and n=1 they
coincide with the Sklyanin algebras. We prove that the Poisson
structure is compatible with the Lie-Poisson structure on the
direct sum of n copies of sl(N). The derivation is based on the
Poisson reduction from the canonical brackets on the affine space
over the cotangent bundle to the groups of automorphisms of vector
bundles.
\end{abstract}
\today

\section{Introduction}

In this article we construct  quadratic Poisson algebras (the
classical Sklyanin-Feigin-Odesskii algebras) based on the exchange
relations with the Belavin-Drinfeld elliptic $\sln$ $r$-matrix
\cite{BD} and their quantum version related to the vertex elliptic
matrix \cite{B}. These algebras are parameterized by the moduli
space of complex structures of elliptic curves with $n$ marked
points and the Planck constant living on curves in the quantum
case. For SL$_2$ and $n=1$ we come to the original Sklyanin
algebra \cite{Skl}. The constructed algebras are particular case
of general construction \cite{FO}, but in contrast of the generic
case they are finitely generated. We describe explicitly the
Poisson brackets between the generators and the corresponding
quadratic relations in the quantum case in terms of quasi-periodic
functions on the moduli space. In the classical case the Poisson
algebras
 have a form of quadratic algebras on the
direct product of $n$ copies of $\GLN$ with a nontrivial mixing of
the components. On the other hand, there exists the standard
linear Lie-Poisson structure on direct sum $\oplus_{a=1}^n$
Lie$(\GLN)$. We prove that the both Poisson structures are
compatible.

The classical algebras define symmetries  of the elliptic
generalization of the Schlesinger and the Garnier systems
\cite{Ta,CLOZ}.

 In Section 2 we derive the classical vertex r-matrix
and the $\GLN$-valued Lax matrix with n simple poles from the
canonical brackets on some generalization of the cotangent bundle
of the $\GLN$ two-loop group by the Poisson reduction. In section
3 we present the explicit form of the brackets and prove that they
are compatible with the Lie-Poisson brackets. The section 4 is
devoted to the quantum case.

\bigskip
\addcontentsline{toc}{section}{\numberline{}Acknowledgments} The
work  was supported in part by grants RFBR-06-02-17381,
NSch-8065-2006.2 and RFBR-06-01-92054-KE. The work of A.Z. was
also supported by the "Dynasty" fund. A.L and A.Z. are grateful to
the Max Planck Institute of Mathematics at Bonn for the
hospitality where this work was prepared.

\section{Classical exchange relations from $\GLN$ two-loop group}
\setcounter{equation}{0}

\subsection{Degree one vector bundles over elliptic curves}

Let $\Si_\tau=\mC/(\mZ+\tau\mZ)$  be an elliptic curve, with the
modular parameter $\tau$, $(\Im m\tau>0)$.
 Consider a vector bundle $E_N$ of a rank $N$
  over $\Si_\tau$. It is described by its sections\\
$s=(s_1(z,\bz),\ldots,s_N(z,\bz))$   with monodromies
$$
s^T(z+1,\bz+1)=Qs^T(z,\bz)\,,~~s^T(z+\tau,\bz+\bar{\tau})=\ti{\La}s^T(z,\bz)\,,
$$
where
$$
Q=\di(1,\bfe_N,\ldots,\bfe_N^{N-1})\,,~
\bfe_N=\exp\frac{2\pi\imath}{N}\,,~
\ti{\La}=\bfe_N^{-(z+\frac{\tau}{2})}\La\,,\,~\La=(E_{j,j+1})\,,
$$
where $E_{j,j+1}$ is a matrix with a unity on the $(j,j+1)$ place.
Since $\det Q=\pm 1$ and $\det\ti{\La}=\pm
\bfe_1^{-(z+\frac{\tau}{2})}$ the determinants of the transition
matrices have the same quasi-periods as the Jacobi
theta-functions. The  theta-functions have simple poles in the
fundamental domain $\Si_\tau$. Thereby, the vector bundle $E_N$
has degree one.

One can choose a holomorphic section $(\bp s=0)$ in the form
$$
s(z)=(\thch{\frac{1}{N}}{0}{z}{N\tau},\ldots,\thch{1}{0}{z}{N\tau})\,.
$$
Define the transformations
$s^T\to f(z,\bz)s^T$ by smooth maps $f\,:\,\Si_\tau\to\GLN$\\
$(f\in\Om^{(0,0)}_{C^\infty}(\Si_\tau,\GLN))\,,$ with monodromies
$$
f(z+1,\bz+1)=Q^{-1}f(z,\bz)Q\,,~~f(z+\tau,\bz+\bar{\tau})
=\ti{\La}^{-1}f(z,\bz)\ti{\La}\,.
$$
They preserve the degree of $E_N$ and therefore generate the gauge
group $\mathcal{G}=\{f(z,\bz)\}$ of $E_N$.

In general, the operators
$$
d_{\bA}=\bp+\bA\,:\, \Om^{(0,0)}(\Si_\tau,
E_N)\to\Om^{(0,1)}(\Si_\tau,E_N)
$$
 define a complex structure of $E_N$. A section is holomorphic if
$d_{\bA}(s^T)=0$. Here we assume that $\bA$ has the same
monodromies as the sections of $E_N$ \beq{AM}
\bA(z+1,\bz+1)=Q^{-1}\bA(z,\bz)Q\,,~~\bA(z+\tau,\bz+\bar{\tau})
=\ti{\La}^{-1}\bA(z,\bz)\ti{\La}\,. \eq Two complex structures,
defined by $\bA$ and $\bA^f$, are called equivalent if they are
related by the gauge transform \beq{gt1} \bA^f=f^{-1}\bA
f+f^{-1}\bp f\,,~~f\in\mathcal{G} \,. \eq
 The quotient of the space of generic connections
$\mathcal{A}=\{\bA\}$ with respect to the $\mathcal{G}$-action is
the moduli space of holomorphic bundles
$Bun(E_N)=\mathcal{A}/\mathcal{G}$.

Consider  the two-loop group $LL(\GLN)$ represented by the space
of sections\\
 $\{g(z,\bz)\}=\Om^{(0,0)}_{C^\infty}(\Si_\tau,\GLN)$
with the monodromies \beq{gM} g(z+1,\bz+1)=Q^{-1}g(z,\bz)Q\,,~~
g(z+\tau,\bz+\bar{\tau})=\ti{\La}^{-1}g(z,\bz)\ti{\La}\,. \eq The
two-loop group $LL(\GLN)$ with these quasi-periodicity conditions
is the group $Aut E_N$ of automorphisms of the degree 1 vector
bundle $E_N$.

\subsection{The Poisson structure on $\clR$}

The space
$\mathcal{R}=\mathcal{A}\times\Om^{(0,0)}_{C^\infty}(\Si_\tau,\GLN)
=\{(\bp+\bA,g)\}$ can be endowed with the symplectic form \beq{sf}
 \om=
\int_{\Si_\tau}K \lan d(\bA g^{-1})\wedge
dg\ran+\oh\int_{\Si_\tau}K\lan g^{-1}dg\wedge \bp(g^{-1}dg)\ran\,,
\eq where $\lan\,,\,\ran$ is the trace in the vector
representation, and $K$ is a section of the canonical bundle over
$\Si_\tau$ $(K\in\Om^{(1,0)}(\Si_\tau))$. We choose $K=dz$. The
space $\mathcal{R}$ is the affine space over the cotangent bundle
to the two-loop group $T^*(LL(\GLN))$.

The transformations (\ref{gt1}) along with \beq{gt2} g\to f^{-1}gf
\eq are canonical with respect to the symplectic form (\ref{sf}).
The Hamiltonian vector fields $V_\ep$, $(\ep\in Lie(\mathcal{G}))$
on $\mathcal{R}$ $(V_\ep\om=d\mu^*)$ are generated by the
Hamiltonian \beq{mm} \mu^*(\ep;\bA,g)=\int_{\Si_\tau}K
\lan\ep(g\bA g^{-1}-\bp gg^{-1}-\bA)\ran\,. \eq
\begin{rem}
Let $\Phi\in\Om^{(1,0)}_{C^\infty}(\Si_\tau,End E_N)$ be the Higgs
field and $g=\exp(\hbar K^{-1}\Phi)$, where $\hbar\in\mC$. In the
limit $\hbar\to 0$ \beq{lim}
 g\sim K^{-1}(Id+\hbar\Phi+\ldots)\,.
\eq The form (\ref{sf}) in the first order becomes the canonical
form on the Higgs bundle $\{(d_{\bA},\Phi)\}$. The symmetries
defines the Hamiltonian
$\mu^*(\ep;\bA,\Phi)=\int_{\Si_\tau}\lan\ep(\bp\Phi+[\bA,\Phi])\ran$.
Thus,   $\mathcal{R}$ is a deformation of the Higgs bundle.
\end{rem}

 The inversion of (\ref{sf})
defines the Poisson structure on $\mathcal{R}$. In terms of
coordinates in the basis (\ref{B.11})\\
 $\bA=\sum_{\al\in\ti{\mZ}^{(2)}_N}\bA_\al T_\al$ and
$g=\sum_{a\in \mZ^{(2)}_N}g_aT_a$ it takes the form \beq{pb1}
K\{\bA_\al(z,\bz),\bA_\be(w,\bw)\}=
C_{\al+\be}\bA_{\al+\be}\de(z-w,\bz-\bw)+\bp\de(z-w,\bz-\bw)\de_{\al,-\be}\,,
\eq \beq{pb2} K\{g_a(z,\bz),\bA_\be(w,\bw)\}= \bfe_{N}^{(a\times
\be)}g_{a+\be}(z,\bz)\de(z-w,\bz-\bw)\,,
 \eq
 \beq{pb3}
\{g_a(z,\bz),g_b(w,\bw)\}=0\,. \eq The brackets define the Poisson
algebra  $\clO(\mathcal{R})$
 with the symmetry group $\mathcal{G}$.

Define a Poisson subalgebra $\bfP_{\Si_\tau}$ of
$\clO(\mathcal{R})$. It satisfies the following conditions:
\begin{enumerate}
  \item The connection $\bA$ takes values in the
subalgebra $\sln$, while the field $g$ is still takes value in
$\GLN$;
  \item $\bfP_{\Si_\tau}$ is generated by holomorphic functionals over $\clR$
  with the test functions vanishing at $z=0$.
\end{enumerate}

 The subalgebra $\bfP_{\Si_\tau}$ has a center
$\mathcal{Z}$ generated by  $\det g(z,\bz)$. The symmetry group
$\mathcal{G}^s\subset\mathcal{G}$ of $\bfP_{\Si_\tau}$ is
generated by the smooth maps $f\,:\,\Si_\tau\to\SLN$.


\subsection{The Poisson reduction}

 Our goal is calculating the reduced Poisson structure with respect to the
$\mathcal{G}^s$-action. The standard Poisson reduction
$\bfP^{red}_{\Si_\tau}$ of $\bfP_{\Si_\tau}$ is described  as
follows.  Let $\bfP_{\Si_\tau}^{\mathcal{G}^s}$ is the invariant
Poisson subalgebra and
$$
I^{\mathcal{G}^s}=\{\mu^*(\ep)F(\bA,g)\,|\,F(\bA,g)\in\bfP_{\Si_\tau}\}
$$
is the ideal in $\bfP_{\Si_\tau}^{\mathcal{G}^s}$ generated by the
functional $\mu^*(\ep)$ (\ref{mm}), where $\ep\in
Lie(\mathcal{G}^s)$. The reduced Poisson algebra
$\bfP^{red}_{\Si_\tau}$ is  the factor algebra \beq{fs}
\bfP^{red}_{\Si_\tau}=\bfP_{\Si_\tau}^{\mathcal{G}^s}/I^{\mathcal{G}^s}
:=\bfP_{\Si_\tau}//\mathcal{G}^s\,. \eq In our construction we use
another ideal in $\bfP_{\Si_\tau}^{\mathcal{G}^s}$. It will be
defined below.

First, calculate the brackets in the invariant subalgebra
$\bfP_{\Si_\tau}^{\mathcal{G}^s}$. Due to the monodromy conditions
(\ref{AM}) the generic field $\bA$ is gauge equivalent to the
trivial $f^{-1}\bA f+f^{-1}\bp f=0$. Therefore \beq{pg} \bA=-\bp
f[\bA]f^{-1}[\bA]\,. \eq
Again, the monodromies of the gauge
matrices (\ref{gM}) prevent to have nontrivial residual gauge
symmetries. Let $f[\bA](z,\bz)$ be a solution of (\ref{pg}).
Consider the transformation of $g$ by solutions of (\ref{pg})
\beq{lr}
L[\bA,g](z,\bz)=f[\bA](z,\bz)g(z,\bz)f^{-1}[\bA](z,\bz)\,. \eq The
gauge invariant subalgebra $\bfP_{\Si_\tau}^{\mathcal{G}^s}$ is
generated by the matrices $L$
$$
\bfP_{\Si_\tau}^{\mathcal{G}^s}=\{\Psi(\bA,g)=\Psi(0,L)\}
$$

\begin{predl}
The brackets on $\bfP_{\Si_\tau}^{\mathcal{G}^s}$ take the form of
the classical exchange relations \beq{qb}
\{L_1(z,\bz),L_2(w,\bw)\}=[r(z-w),L_1(z,\bz)\otimes L_2(w,\bw)]\,,
\eq
 where $L_1(z,\bz)=L(z,\bz)\otimes Id,\,$ $L_2(w,\bw)=Id\otimes
L(w,\bw)$, and $r(z,w)$ is the classical Belavin-Drinfeld elliptic
$r$-matrix \cite{BD}.
\end{predl}
{\it Proof}.\\
The calculation of brackets in $\bfP_{\Si_\tau}^{\mathcal{G}^s}$
is reduced to the calculation on shell ($\bA=0$, $f=Id$) of the
Poisson brackets between the matrix elements of (\ref{lr}) by
(\ref{pb1}) - (\ref{pb3}). In doing these calculations we need
only the expression \beq{rexp}
r_{\al,\be}(z,\bz;z',\bz')=\frac{\de
f_\al(z,\bz)}{\de\bA_\be(z',\bz')}|_{\bA=0}\,. \eq
 The
straightforward  calculations of the brackets $\{L,L\}$ performed
in \cite{BDOZ} lead to the desired r-matrix form (\ref{qb}).

Let us find the  r-matrix. Due to (\ref{pg}) $r$ is the Green
function of the operator $\bp$ \beq{gf}
 \bp
r_{\al,\be}(z,\bz;z',\bz')=\de_{\al+\be,0}\de(z-z',\bz-\bz')\,,
\eq having the following quasi-periodicities
$$
r(z+1,\bz+1)=(Q^{-1}\otimes Id)\, r(z,\bz)\,(Q\otimes Id)\,,
$$
$$
 r(z+\tau,\bz+\bar{\tau})=(\La^{-1}\otimes Id)\, r(z,\bz)\,(\La\otimes Id)\,.
$$
It follows from (\ref{gf}) that $r_{\al,\be}$ is a meromorphic and
singular  on the diagonal \beq{di} \lim_{z'\to
z}r_{\al,\be}(z,z')=\f1{z-z'}T_\al\otimes T_\be\de_{\al+\be,0}\,.
\eq Due to (\ref{A.3a} ), (\ref{vf} ), and (\ref{qpe1} )
\beq{rmat} r(z,w)=r(z-w)=\sum_\al \varphi_\al(z-w)T_\al\otimes
T_{-\al}\,. \eq It is the Belavin-Drinfeld classical r-matrix
\cite{BD}. This $r$-matrix satisfies the classical Yang-Baxter
equation providing the Jacoby identity for the brackets
(\ref{qb}).  $\Box$

\begin{rem}
In the limit (\ref{lim}) the only non-trivial brackets (\ref{pb2})
assume the form
$$
\{\Phi_\al(z,\bz),\bA_\be(w,\bw)\}= \de_{\al,-\be}\de(z-w,\bz-\bw)
$$
and (\ref{qb}) is replaced by the linear brackets
$$
\{L_1(z,\bz),L_2(w,\bw)\}=[r(z-w),L(z,\bz)\otimes Id+ Id\otimes
IdL_2(w,\bw)]\,.
$$
\end{rem}

Let us fix  a divisor  of non-coincident  points on $\Si_\tau$
$$
D_n=(x_1,\ldots,x_n)\,,~x_j\neq x_k\,,~x_j\in\Si_\tau\,.
$$
Define the subalgebra $Lie(D_n)(\mathcal{G}^s)\subset
Lie(\mathcal{G}^s)$
$$
Lie(D_n)(\mathcal{G}^s)=\{\ve\in
Lie(\mathcal{G}^s)\,|\,\ve(x_j,\bar{x}_j)=0\,,~x_j\in D_n\}\,.
$$
Consider the ideal $I(D_n)$ generated by the functional \beq{cs}
\mu_{D_n}^*(\ve;\bA,g)
=\mu_{D_n}^*(\ve;L)=\int_{\Si_\tau}\lan\ve\bp L(z,\bz)\ran\,, \eq
where $\ve$ belongs to $Lie(\mathcal{G}^s(D_n))$. Since
$\mu_{D_n}^*$ depends only on $L\,$,
$I(D_n)\subset\bfP_{\Si_\tau}^{\mathcal{G}^s}$.

\bigskip

Consider the quotient Poisson algebra
$\bfP_{\Si_\tau}^{\mathcal{G}^s}/I(D_n)$
\begin{predl}
 The reduced Poisson algebra
\beq{fs1}
\bfP_{\Si_\tau,D_n}^{red}=\bfP_{\Si_\tau}^{\mathcal{G}^s}/I(D_n)\,,
\eq is finitely generated
$$
\dim \bfP_{\Si_\tau,D_n}^{red}=nN^2\,.
$$
The matrix $L(z)$ in the classical exchange relations ($\emph{the
Lax matrix}$) takes the form \beq{lax} L = S_{0}T_{0} +
\sum_{j=1}^{n}(S_{0}^{j}E_{1}(z-x_{j})T_{0} + \tilde{L}_{j})\,, \
\ \ \tilde{L}_{j} =
\sum_{\alpha}S_{\alpha}^{j}\varphi_{\alpha}(z-x_{j})T_{\alpha}\,,
\eq where \beq{res} \sum_{j=1}^nS_{0}^{j}=0\,,
 \eq
 $\varphi_{\alpha}(z-x_{j})$ are defined by (\ref{vf}), and
 $E_{1}(z-x_{j})$ is the first Eisenstein series (\ref{A.1}).
\end{predl}
{\it Proof}\\
  To prove it we analyze solutions of  (\ref{cs}):
\beq{cs1} \mu_{D_n}^*(\ve;L)=\int_{\Si_\tau}\lan\ve\bp
L(z,\bz)\ran\ =0 \,. \eq The solutions are meromorphic
quasi-periodic maps having simple poles at the marked points. Let
$L(z)=\sum_{a}L_a(z)T_a$ be the expansion of $L$ in the basis
$T_a$ of $\GLN$. It follows from (\ref{gM}) that
$$
L_a(z+1)=\bfe_N^{a_2}L_a(z)\,,~~L_a(z+\tau)=\bfe_N^{-a_1}L_a(z)\,,~~a=(a_1,a_2)\,.
$$
The functions $\varphi_a(z-x_j)$ (\ref{vf}) have these monodromies
(\ref{qpe1}) and simple poles at $x_j$. They form a
$n$-dimensional basis in the space of quasi-periodic functions
(\ref{qp}) with the poles at $x_j$.
   If $a=(0,0)$ then
$L_0(z)$ is a double-periodic function with simple poles at $x_j$.
The basis in this space is $1$ and the Eisenstein functions
$E_{1}(z-x_{j})$ with vanishing sum of their residues. Thus, the
space has dimension $n$.
 In this way we come to (\ref{lax}) and (\ref{res}). $\Box$

 \bigskip

As we mentioned above $\det g$ generates the Casimir functionals
in $\bfP^{red}_{\Si_\tau}$. Thereby, the brackets on
$\bfP^{red}_{\Si_\tau,D_n}$ are degenerate. The function $\det
L(z)$ is the generating function for the Casimir elements
$C^\mu(j)$. Since $\det L(z)$ is a double periodic function it can
be expanded in the basis of elliptic functions (\ref{A.2a})
\beq{6.10} \det L(z)=C^0+\sum_j^n
C^1(j)E_1(z-x_j)+C^2(j)E_2(z-x_j)+\ldots+C^N(j)E_N(z-x_j)\,. \eq
Due to the condition \beq{7.10} \sum_{j=1}^nC^1(j)=0\,, \eq the
number of the independent Casimir  is $Nn$.
 The generic
symplectic leaf
$$
{\cal
R}^2_{n,N}=\bfP^{red}_{\Si_\tau,D_n}/\{(C^\mu(j)=C(j)^\mu_{(0)})\,,~
\mu=1,\ldots,N\,,j=1,\ldots,N\}\,.
$$
has dimension \beq{10.1} \dim({\cal P}^2_{n,N})=nN(N-1)\,. \eq
Note that it coincides with the sum of dimensions of $n$ generic
$\GLN$ coadjoint orbits.


\section{The structure of the reduced Poisson space}
\setcounter{equation}{0}

\subsection{Explicit form of quadratic brackets}

 Proposition 1.2 provides  the reduced Poisson algebra
$\bfP^{red}_{\Si_\tau,D_n}$ with the generators \beq{gen}
\{S_0\,,\,(S_0^j\,,
\bfS^j=\{S_\al^j\}\,\,,j=1,\ldots,n)\,|\,\sum_{j=1}^nS^j_0=0\}\,.
\eq The brackets between generators were calculated in
\cite{CLOZ}.
\begin{predl}
The Poisson brackets on the space $\mC^{nN^2}$
 in terms of the generators (\ref{gen})
take the form \beq{3.3} \{S_0,S_0^j\}_2
=\{S_0^j,S_0^k\}_2=\{S_\al^j,S_\al^k\}_2=0\,, \footnote{The
subscript index $\{\,,\,\}_2$ means the quadratic brackets.} \eq
\beq{3.4} \{S_0,S_\al^k\}_2= \sum_{\ga\neq\al}\bfC(\al,\ga)\left(
S^k_{\al-\ga}S^k_\ga E_2(\bfga) -\sum_{j\neq
k}S^j_{-\ga}S^k_{\al+\ga}f_\ga(x_k-x_j) \right)\,, \eq \beq{3.5}
\{S^k_\al,S^k_\be\}_2=\bfC(\al,\be)S_0S^k_{\al+\be} +
\sum_{\ga\neq\al,-\be}\bfC(\ga,\al-\be)S^k_{\al-\ga}S^k_{\be+\ga}\bff_{\al,\be,\ga}
\eq
$$
+\bfC(\al,\be)S^k_0S^k_{\al+\be}(E_1(\bfal+\bfbe)-E_1(\bfal)-E_1(\bfbe))
$$
$$
-\bfC(\al,\be)\sum_{j\neq k}
[S_0^kS_{\al+\be}^j\vf_{\al+\be}(x_k-x_j)-
S_0^jS_{\al+\be}^kE_1(x_k-x_j)]\}
$$
$$
-2\sum_{j\neq
k}\bfC(\ga,\al-\be)S^k_{\al-\ga}S^k_{\be+\ga}\vf_{\be+\ga}(x_k-x_j)\}
\,,
$$
where  $\bff_{\al,\be,\ga}$, $E_2(\bfal)$, $E_1(\bfal)$ are
defined by (\ref{fzh}) and (\ref{AA50}). For $j\neq k$ \beq{3.6}
\{S_\al^j,S_\be^k\}_2=\sum_{\ga\neq\al,-\be}
\bfC(\ga,\al-\be)S^j_{\al-\ga}S^k_{\be+\ga}\vf_\ga(x_j-x_k) \eq
$$
-\bfC(\al,\be) \left( S_0^jS_{\al+\be}^k\vf_\al(x_j-x_k)-
S_0^kS_{\al+\be}^j\vf_{-\be}(x_k-x_j) \right)\,,
$$
and \beq{3.7} \{S_0^j,S_\be^k\}_2= \left\{
\begin{array}{cc}
2\sum_\ga \bfC(\ga,-\be)S^j_{-\ga}S^k_{\be+\ga}
\vf_{\ga}(x_k-x_j)\,, & j\neq k \,,\\
-2\sum_{m\neq k}\sum_\ga \bfC(\ga,-\be)S^k_{-\ga}S^m_{\be+\ga}
\vf_{\be+\ga}(x_k-x_m) \,, & j=k\,.
\end{array}
\right. \eq
\end{predl}

This algebra is an explicit particular form of general
construction of quadratic Poisson algebras \cite{FO}. For $n=1$
this algebra was calculated in \cite{HLO} and for $n=1$ and $N=2$
it is the classical Sklyanin algebra \cite{Skl}.


\subsection{Twisting bundles}

We need another but equivalent form of this algebra on
$\mC^{nN^2}$. Consider the twisted bundle\\
$E_N'=Aut(E_N)\otimes\mathcal{L}$, where $\mathcal{L}$ is a
trivial line bundle over $\Si_\tau$. The sections of $E_N'$ are
the sections of $E_N$ multiplied by
$\vartheta(z+\eta)/\vartheta(z)$, $\,(\eta\in\Si_\tau)$.
Therefore, the transition functions of $E_N'$ are
$$
ad(Q)~~{\rm for}~z\to z+1\,,~~~ \exp (-2\pi \imath \eta) \cdot
ad(\ti{\La})~~{\rm for}~z\to z+\tau\,.
$$
It follows from (\ref{vf1}), (\ref{qpe2}) that solutions of
(\ref{cs}) with these monodromies and simple poles at the divisor
$$
\ti{D}_n=(\ti{x_1},\ldots,\ti{x}_n)
$$
 is
\beq{tl1} L^\eta_{\ti{D}_n} = \sum_{j=1}^{n} \left[
\widetilde{S}_{0}^{j}\varphi_{\eta}(z-\widetilde{x}_{j})T_{0} +
\sum_{\alpha}\widetilde{S}_{\alpha}^{j}\varphi_{\alpha,\eta}(z-\widetilde{x}_{j})
T_{\alpha} \right]\,, \eq where $\varphi_{\eta}=\varphi_{0,\eta}$.
The corresponding algebra $\ti{\bfP}^{red}_{\Si_\tau,\ti{D}_n}$ is
defined, as above, by the classical exchange relations \beq{er}
\{L^\eta_{1,\ti{D}_n}(z),L^\eta_{2,\ti{D}_n}(w)\}=
[r(z-w),L^\eta_{1,\ti{D}_n}(z)\otimes L^\eta_{2,\ti{D}_n}(w)]\,.
\eq with the set of $nN^2$ generators
$$
\ti{\bfS}^j=\{\ti{S}^j_a\}\,,~~(a=(a_1,a_2)\in\mZ^{(2)}_N
\,,~j=1,\ldots,n)\,.
$$
The brackets between the generators can be extracted from
(\ref{er}) as before. We do not need their explicit form  because
we prove immediately the equivalence of these two algebras. The
only thing we need in next Section is the brackets containing
$\ti{S}_0^k$ in the rhs (compare with (\ref{3.5}) and (\ref{3.6}))
\beq{exbr} \{\ti{S}_\al^k,\ti{S}_\be^k\}=
\sum_{\ga\neq\al,-\be}\bfC(\ga,\al-\be)\ti{S}^k_{\al-\ga}\ti{S}^k_{\be+\ga}
\ti{\bff}^\eta_{\al,\be,\ga} \eq
$$
+\bfC(\al,\be)\ti{S}^k_0\ti{S}^k_{\al+\be}(E_1(\bfal+\bfbe+\eta)-E_1(\bfal)-
E_1(\bfbe)-E_1(\eta))
$$
$$
-\bfC(\al,\be)\sum_{j\neq k}
[\ti{S}_0^k\ti{S}_{\al+\be}^j\vf_{\al+\be,\eta}(x_k-x_j)-
\ti{S}_0^j\ti{S}_{\al+\be}^k\vf_{\eta}(x_k-x_j)]\}
$$
$$
-2\sum_{j\neq
k}\bfC(\ga,\al-\be)\ti{S}^k_{\al-\ga}\ti{S}^k_{\be+\ga}\vf_{\be+\ga,\eta}(x_k-x_j)\}
\,,
$$
where \beq{tif}
\ti{\bff}^\eta_{\al,\be,\ga}=E_1(\bfga)-E_1(\bfal-\bfbe-\bfga)+
E_1(\bfal-\bfga+\eta)-E_1(\bfbe+\bfga+\eta)\,, \eq \beq{exbr1}
 \{\ti{S}_\al^j,\ti{S}_\be^k\}=\sum_{\ga\neq\al,-\be}
\bfC(\ga,\al-\be)\ti{S}^j_{\al-\ga}\ti{S}^k_{\be+\ga}\vf_{\ga,\eta}(x_j-x_k)
\eq
$$
-\bfC(\al,\be) \left(
\ti{S}_0^j\ti{S}_{\al+\be}^k\vf_{\al,\eta}(x_j-x_k)-
\ti{S}_0^k\ti{S}_{\al+\be}^j\vf_{-\be,\eta}(x_k-x_j) \right)\,.
$$

To prove the equivalence we choose for simplicity  $\ti{x}_i=0$
for some $i$.
\begin{predl}
Fix two indices $1\leq i,k\leq n$ $\,(i\neq k)$. Define
$x_k=-\eta$ and $x_j=\ti{x_j}$ for $j\neq k$. Then Poisson
algebras ${\bfP}^{red}_{\Si_\tau,D_n}$ and
$\ti{\bfP}^{red}_{\Si_\tau,\ti{D}_n}$ are isomorphic. The
corresponding canonical transformations are
$$
S_{0} = \widetilde{S}_{0}^{i} + \sum_{j \neq i}^{n}
\frac{1}{\varphi_{\eta}(\widetilde{x}_{j})}
(E_{1}(\widetilde{x}_{j})+ E_{1}(\eta))\widetilde{S}_{0}^{j}\,,
$$
\beq{tl51} S_{0}^{k} = -\sum_{j \neq i}^{n}
\frac{\widetilde{S}_{0}^{j}}{\varphi_{\eta}(\widetilde{x}_{j})}\,,
\ \ S_{\alpha}^{k} = \frac{\widetilde{S}_{\alpha}^{i}}
{\varphi_{\alpha}(\eta)} + \sum_{j \neq i}
\frac{\varphi_{-\alpha}(\widetilde{x}_{j})\widetilde{S}_{\alpha}^{j}}
{\varphi_{\eta}(\widetilde{x}_{j})\varphi_{\alpha}(\eta)}\,, \eq
$$
S_{0}^{j \neq k} =
\frac{\widetilde{S}_{0}^{j}}{\varphi_{\eta}(\widetilde{x}_{j})}\,,
\ \ \ S_{\alpha}^{j \neq k} =
\frac{\widetilde{S}_{\alpha}^{j}}{\varphi_{\eta}(\widetilde{x}_{j})}\,.
$$
\end{predl}
{\it Proof}\\
The Lax operator $L^\eta_{\ti{D}_n}$ (\ref{tl1}) after dividing on
$\varphi_\eta(z)$ acquires the same monodromies as $L$
(\ref{lax}). Consider the residues and the constant terms of these
operators. First, we have: \beq{tl3} L^\eta_{\ti{D}_n}
/\varphi_{\eta}(z) = \widetilde{S}_{0}^{i}T_{0} + \sum_{j \neq
i}^{n} \left[
\widetilde{S}_{0}^{j}\frac{\varphi_{\eta}(z-\widetilde{x}_{j})}{\varphi_{\eta}(z)}T_{0}+
\sum_{\alpha}\left(\widetilde{S}_{\alpha}^{j}
\frac{\varphi_{\alpha,\eta}(z-\widetilde{x}_{j})}{\varphi_{\eta}(z)}
+
\widetilde{S}_{\alpha}^{i}\frac{\varphi_{\alpha,\eta}(z)}{\varphi_{\eta}(z)}\right)
T_{\alpha} \right]\,. \eq Applying (\ref{ir21}), (\ref{ir22}), and
(\ref{ir23}) we get
$$
L^\eta_{\ti{D}_n}/\varphi_{\eta}(z) = \left(\widetilde{S}_{0}^{i}
+ \sum_{j \neq i}^{n} \frac{1}{\varphi_{\eta}(\widetilde{x}_{j})}
(E_{1}(\widetilde{x}_{j})+ E_{1}(\eta))\widetilde{S}_{0}^{j}
\right) \cdot T_{0}
$$
$$
 - E_{1}(z+\eta) \cdot \sum_{j \neq i}^{n}
\frac{\widetilde{S}_{0}^{j}}{\varphi_{\eta}(\widetilde{x}_{j})}
\cdot T_{0} + \sum_{j \neq i}^{n}E_{1}(z-\widetilde{x}_{j}) \cdot
\frac{\widetilde{S}_{0}^{j}}{\varphi_{\eta}(\widetilde{x}_{j})}
\cdot T_{0}
$$
$$
+\sum_{\alpha,j \neq i} \varphi_{\alpha}(z-\widetilde{x}_{j})
\frac{\widetilde{S}_{\alpha}^{j}}{\varphi_{\eta}(\widetilde{x}_{j})}
\cdot T_{\alpha} + \sum_{\alpha,j \neq i} \varphi_{\alpha}(z+\eta)
\cdot \left(
\frac{\varphi_{-\alpha}(\widetilde{x}_{j})\widetilde{S}_{\alpha}^{j}}
{\varphi_{\eta}(\widetilde{x}_{j})\varphi_{\alpha}(\eta)} +
\frac{\widetilde{S}_{\alpha}^{i}} {\varphi_{\alpha}(\eta)} \right)
\cdot T_{\alpha}
$$
Note that there is a new pole at $x_{b}= -\eta$. Comparing with
(\ref{lax}) we come to (\ref{tl51}). $\Box$


\subsection{Bihamiltonian structure}

Introduce on the space $\mC^{nN^2}$ the linear (Lie-Poisson)
brackets. To this end consider the direct sum of $n$ copies of
$\gln\,$: $\,\gg^*=\gln\oplus\ldots\oplus\gln$ with the brackets
\beq{lin}
 \{S^j_{\alpha},S^k_{\beta} \}_{1}= C(\alpha,\beta) S^j_{\alpha
+ \beta}\de^{jk} \,. \footnote{The subscript index $_1$ means the
linear brackets.}
 \eq
\begin{rem}
The Lie-Poisson brackets have the $r$-matrix form
$$
\{\ti{L}_1(z),\ti{L}_2(w)\}=[r(z-w),\ti{L}_1(z)+\ti{L}_2(w)]\,,
$$
where $r$ is same as for the quadratic brackets (\ref{rmat}), and
$\ti{L}=\sum_{j=1}^n\ti{L_j}(z)$ (\ref{lax}).
\end{rem}

Two Poisson structures are called \textit{compatible} (or, form
\textit{Poisson pair}) if their linear combinations are Poisson
structures as well.

\begin{predl}
The linear (\ref{lin}) and quadratic (\ref{3.3}) - (\ref{3.7})
Poisson brackets on the space $\mC^{nN^2}$ are compatible.
\end{predl}
\textit{Proof.}\\
Choose a point $x_k\in \ti{D}_n$ and replace the  variable
$\ti{S}_{0}^{k}$ by $\ti{S}_{0}^{k} + \lambda$, where $\la\in\mC$
is a number and therefore it Poisson commutes with all elements of
the quadratic Poisson algebra. Substitute the new variable in
(\ref{exbr}) and (\ref{exbr1}).
 The change of variables does not spoil the Jacobi identity and therefore
  we come to the following Poisson structure
$$
\{ \ti{S},\ti{S}\}_{\lambda} := \{ \ti{S},\ti{S} \}_{2} + \lambda
\{ \ti{S},\ti{S}\}_{1}\,.
$$
Consider the linear brackets term. \beq{bih3}
\begin{array}{c}
 \{\ti{S}^{k}_{\alpha},\ti{S}^{k}_{\beta} \}_1 =
F_{1}\ti{S}^{j}_{\alpha+\beta} + F_{2}\ti{S}^{k}_{\alpha+\beta},\\
\{ \ti{S}^{k}_{\alpha},\ti{S}^{j}_{\beta} \}_1 =
G_{2}\ti{S}^{j}_{\alpha+\beta},\\
\{ \ti{S}^{j}_{\alpha},\ti{S}^{k}_{\beta} \}_1 =
\tilde{G}_{2}\ti{S}^{j}_{\alpha+\beta},\\
\{ \ti{S}^{j}_{\alpha},\ti{S}^{j}_{\beta} \}_1 =
H_{2}\ti{S}^{j}_{\alpha+\beta}\,,
\end{array}
\eq where up to the common multiplier $C(\al,\be)$ the
coefficients have the form \beq{bih4}
\begin{array}{c}
F_{1}= \varphi_{\alpha + \beta,\eta}(x_{kj}),\ \ \ F_{2}=
-E_{1}(\bfal) - E_{1}(\bfbe) - E_{1}(\eta) + E_{1}(\bfal + \bfbe + \eta),\\
G_{2}= -\varphi_{\alpha}(x_{kj}),\ \ \
\tilde{G_{2}}= -\varphi_{\beta}(x_{kj}),\\
H_{2}=\varphi_{0,-\eta}(x_{kj})\,,
\end{array}
\eq where $x_{kj}=x_{k}-x_{j}$ The following Lemma completes the
proof.
\begin{lem}
The linear Poisson algebra (\ref{bih3}) is equivalent to the
direct sum of Lie-Poisson algebras on $\oplus_{l=1}^n\gln$
\beq{nv} \{t^j_\al,t^k_\be\}=C(\al,\be)t^j_{\al+\be}\de^{jk}\,.
\eq
\end{lem}
{\it Proof}\\
Define
$$
\ti{S}^{k}_{\alpha}=a_\al t^k_\al+b_\al t^j_\al\,,~~
\ti{S}^{j}_{\alpha}=H_2t^j_\al
$$
The brackets (\ref{bih3}) are equivalent to (\ref{nv}) if
\beq{bih7}
\begin{array}{c}
a_{\alpha}a_{\beta} = a_{\alpha + \beta} F_{2},\\
b_{\alpha} = G_{2},\\
b_{\alpha}b_{\beta} = F_{1}H_{2} + b_{\alpha + \beta}F_{2}.
\end{array}
\eq Let us solve these equations. The solution of the first
equation can be found from (\ref{i}). It takes the form
$a_\al=-\varphi_\al(\eta)$. Next prove that
$b_\al=G_2=-\varphi_\al(x_{kj})$ satisfies the last relation. With
$b_\al=-\varphi_\al(x_{kj})$ it takes the form
$$
\varphi_{\alpha}(x_{kj})\varphi_{\beta}(x_{kj})= \varphi_{\alpha
+\beta,\eta}(x_{kj})\varphi_{0,-\eta}(x_{kj})+
$$
$$
+\varphi_{\alpha + \beta}(x_{kj}) \left(  E_{1}(\bfal) +
E_{1}(\bfbe) + E_{1}(\eta) - E_{1}(\bfal + \bfbe +\eta) \right)\,.
$$
It follows from (\ref{ad31}) that
$$
\varphi_{\alpha +\beta,\eta}(x_{kj})\varphi_{0,-\eta}(x_{kj})=
\varphi_{\alpha +\beta}(x_{kj}) \left(  E_{1}(\bfal + \bfbe
+\eta)) + E_{1}(-\eta) + E_{1}(x_{kj}) - E_{1}(\bfal + \bfbe +
x_{kj}) \right)\,,
$$
so the last relation in (\ref{bih7}) is an identity and thereby we
come from (\ref{bih3}) to (\ref{nv}). $\Box$


\section{Quantum algebra}
\subsection{General case}
\setcounter{equation}{0}

In this section we consider quantization of quadratic Poisson
algebra for the case $n>1$. Let us consider quantum $R$-matrix,
having the following form:

\beq{qc1} R(z,w)= \sum_{a \in \mathbb{Z}^{(2)}_{N}}
\varphi_{a}^{\hbar}(z-w)T_{a} \otimes T_{-a}\,, \eq
where we put $\varphi_{a}^{\hbar}(z) \equiv \varphi_{\hbar,
a}(z)$. Note, that in contrast with the classical $r$-matrix,
there is an additional term
$$\varphi_{0}^{\hbar}(z-w) \sigma_{0} \otimes \sigma_{0}.$$

Quantum $R$-matrix satisfies the quantum Yang-Baxter equation:

\beq{qc3}
R_{12}(z-w)R_{13}(z)R_{23}(w)=R_{23}(w)R_{13}(z)R_{12}(z-w)\,. \eq

The quantum Yang-Baxter equation allows us to define the
associative algebra by the relation:

\beq{qc4}
R(z-w)L_{1}^{\hbar}(z)L_{2}^{\hbar}(w)=L_{2}^{\hbar}(w)L_{1}^{\hbar}(z)R(z-w)\,,
\eq

The Lax operator in (\ref{qc4}) has the following monodromies with
respect to $\hbar$:

\beq{qc41} L^{\hbar+\tau}(z)= {\bf e}_N(-z) L^{\hbar}(z)\,, \ \ \
L^{\hbar+1}(z)= L^{\hbar}(z)\,. \eq

So, we have to suppose that the variables $S$ depend on $\hbar$
and $x_{j}$. The new variables and the Lax operator in (\ref{qc4})
takes the following form:

$$
S_{new}^{j} = \hat{S}_{0}^{j}\varphi_{0}^{\hbar}(x_{j})\,,
$$
\beq{qc5}
 L^{\hbar}(z)
= \sum_{j=1}^{n}
\left(\hat{S}_{0}^{j}\varphi_{0}^{\hbar}(x_{j})\varphi_{0}^{\hbar}(z-x_{j})T_{0}
+
\sum_{\alpha}\hat{S}_{\alpha}^{j}\varphi_{\alpha}^{\hbar}(x_{j})\varphi_{\alpha}^{\hbar}(z-x_{j})T_{\alpha}
\right)=
 \eq
$$
=\sum_{j=1}^{n} \sum_{a \in
\mathbb{Z}^{(2)}_{N}}\hat{S}_{a}^{j}\varphi_{a}^{\hbar}(x_{j})\varphi_{a}^{\hbar}(z-x_{j})T_{a}
 \,.
$$

\ \\

\textbf{Proposition 5.1}: \textit{The relations in the associative
algebra assume the form}

$$
\sum_{c}  f^{\hbar}(a,b,c) \cdot
\hat{S}^{j}_{b+c}\hat{S}^{j}_{a-c}\varphi_{b+c}^{\hbar}(x_{j})
\varphi_{a-c}^{\hbar}(x_{j}) \bfe_N(+\frac{c
\times (a-b)}{2}) \ +\\
$$
\beq{qc6}
 + \sum_{c}\sum_{k \neq j}
\varphi_{a-c}^{\hbar}(x_{j}-x_{k})\varphi_{b+c}^{\hbar}(x_{j})
\varphi_{a-c}^{\hbar}(x_{k}) \cdot \eq
$$
\cdot \left( \hat{S}^{j}_{b+c}\hat{S}^{k}_{a-c} \bfe_N(+\frac{c
\times (a-b)}{2}) - \hat{S}^{k}_{a-c}\hat{S}^{j}_{b+c}
\bfe_N(-\frac{c \times (a-b)}{2}) \right) =0,
$$

 \textit{and}:

\beq{qc7} \sum_{c} \varphi_{c}^{\hbar}(x_{j}-x_{k})
\varphi_{b+c}^{\hbar}(x_{k}) \varphi_{a-c}^{\hbar}(x_{j}) \cdot
\left(\hat{S}^{j}_{a-c}\hat{S}^{k}_{b+c} \bfe_N(-\frac{c \times
(a-b)}{2}) - \hat{S}^{k}_{b+c} \hat{S}^{j}_{a-c} \bfe_N(+\frac{c
\times (a-b)}{2}) \right) =0\,, \eq
$$
 k \neq j\,,
$$

\textit{where}
$$f^{\hbar}(a,b,c) =
E_{1}(c+\hbar)-E_{1}(a-b-c+\hbar)
+E_{1}(a-c+\hbar)-E_{1}(b+c+\hbar)
$$
  \textit{and} $a,b,c \in \mathbb{Z}^{(2)}_{N}$.\\

\textit{Proof.}

Let us consider the certain matrix element $T_{a} \otimes T_{b}$.
For this put (\ref{qc5}) and (\ref{qc1}) in (\ref{qc4}), we get
the following expressions:

\beq{qc9}
  \sum_{j,k}\sum_{c,a,b} \varphi_{c}^{\hbar}(z-w)
\varphi_{a}^{\hbar}(z-x_{j})\varphi_{b}^{\hbar}(w-x_{k}) \cdot
\hat{S}_{a}^{j}\hat{S}_{b}^{k} \varphi_{b}^{\hbar}(x_{k})
\varphi_{a}^{\hbar}(x_{j}) \cdot T_{c}T_{a} \otimes T_{-c}T_{b} =
 \eq
$$
= \sum_{j,k}\sum_{c,a,b} \varphi_{c}^{\hbar}(z-w)
\varphi_{a}^{\hbar}(z-x_{j})\varphi_{b}^{\hbar}(w-x_{k}) \cdot
\hat{S}_{b}^{k}\hat{S}_{a}^{j} \varphi_{b}^{\hbar}(x_{k})
\varphi_{a}^{\hbar}(x_{j}) \cdot T_{a}T_{c} \otimes T_{b}T_{-c}\,.
$$

\beq{qc91}
  \sum_{j,k}\sum_{c,a,b} \varphi_{c}^{\hbar}(z-w)
\varphi_{a}^{\hbar}(z-x_{j})\varphi_{b}^{\hbar}(w-x_{k}) \cdot
\hat{S}_{a}^{j}\hat{S}_{b}^{k} \varphi_{b}^{\hbar}(x_{k})
\varphi_{a}^{\hbar}(x_{j}) \bfe_N(-\frac{c \times (a-b)}{2}) \cdot
T_{c+a} \otimes T_{-c+b} =
 \eq
$$
= \sum_{j,k}\sum_{c,a,b} \varphi_{c}^{\hbar}(z-w)
\varphi_{a}^{\hbar}(z-x_{j})\varphi_{b}^{\hbar}(w-x_{k}) \cdot
\hat{S}_{b}^{k}\hat{S}_{a}^{j} \varphi_{b}^{\hbar}(x_{k})
\varphi_{a}^{\hbar}(x_{j}) \bfe_N(+\frac{c \times (a-b)}{2}) \cdot
T_{a+c} \otimes T_{b-c}\,.
$$

The functions of l.h.s and r.h.s. are equal because their poles
and quasi-periods coincide. After changing the variables $a
\rightarrow a - c, \ b \rightarrow b + c$, we get for the
coefficients in front of the matrix element $T_{a} \otimes T_{b}$:

\beq{qc10}
  \sum_{c} \varphi_{c}^{\hbar}(z-w)
\varphi_{a-c}^{\hbar}(z-x_{j})\varphi_{b+c}^{\hbar}(w-x_{k})
\varphi_{b+c}^{\hbar}(x_{k}) \varphi_{a-c}^{\hbar}(x_{j}) \cdot
 \eq
$$
\cdot \left( \hat{S}^{j}_{a-c}\hat{S}^{k}_{b+c} \bfe_N(-\frac{c
\times (a-b)}{2}) - \hat{S}^{k}_{b+c} \hat{S}^{j}_{a-c}
\bfe_N(+\frac{c \times (a-b)}{2}) \right) = 0\,.
$$

We have to consider two types of these expressions:

$$
k \neq j :\\
  \sum_{c} \varphi_{c}^{\hbar}(z-w)
\varphi_{a-c}^{\hbar}(z-x_{j})\varphi_{b+c}^{\hbar}(w-x_{k})
\varphi_{b+c}^{\hbar}(x_{k}) \varphi_{a-c}^{\hbar}(x_{j}) \cdot
$$
$$
\cdot \left( \hat{S}^{j}_{a-c}\hat{S}^{k}_{b+c} \bfe_N(-\frac{c
\times (a-b)}{2}) - \hat{S}^{k}_{b+c} \hat{S}^{j}_{a-c}
\bfe_N(+\frac{c \times (a-b)}{2}) \right) = 0\,,
$$
\beq{qc11} k = j :
 \eq
$$
  \sum_{c} \left( \varphi_{c}^{\hbar}(z-w)
\varphi_{a-c}^{\hbar}(z-x_{j})\varphi_{b+c}^{\hbar}(w-x_{j}) -
\varphi_{a-b-c}^{\hbar}(z-w)
\varphi_{a-c}^{\hbar}(z-x_{j})\varphi_{b+c}^{\hbar}(w-x_{k})
\right) \cdot
$$
$$
\cdot
\hat{S}^{j}_{a-c}\hat{S}^{k}_{b+c}\varphi_{b+c}^{\hbar}(x_{k})
\varphi_{a-c}^{\hbar}(x_{j}) \bfe_N(-\frac{c \times (a-b)}{2}) =
0\,.
$$

We get second expression after changing $c \rightarrow a-b-c$.
Taking the limits ($z \rightarrow x_{j}, w \rightarrow x_{j}$) and
($z \rightarrow x_{j}, w \rightarrow x_{k}$), as it has been
already done in section three, we get the coefficients which must
be equal to zero. So we come to (\ref{qc6}) and (\ref{qc7}).
$\Box$

\subsection{Quadratic algebra in ${\rm GL}(2,{\mathbb C})$ case}

Let us consider the case $N=2$ in more detail. In this case
quantum $R$-matrix take the following form:

\beq{qc22} R(z,w)= \sum_{a =0}^{3} \varphi_{a}^{\hbar}(z-w)
\sigma_{a} \otimes \sigma_{a}, \eq where instead of $T_{a}$ we use
the
basis of sigma-matrices.\\
\\

\textbf{Proposition 5.2}: \textit{The relations in the associative
algebra assume the form}

\beq{qc26} [\hat{S}_{\alpha}^{j}, \hat{S}_{\beta}^{j}]_{-} = i
\varepsilon_{\alpha \beta \gamma}  c_{1}^1(j,j; \al,\be,\ga)
[\hat{S}_{\gamma}^{j}, \hat{S}_{0}^{j}]_{+} +
 \eq
$$
+ \sum_{k \neq j} i \varepsilon_{\alpha \beta \gamma}
\frac{1}{k_{\alpha}} \left(
\varphi_{\gamma}^{\hbar}(x_{jk})c_{1}^2(j,k; \al,\be,\ga)
[\hat{S}_{\gamma}^{k}, \hat{S}_{0}^{j}]_{+} -
\varphi_{0}^{\hbar}(x_{jk})c_{1}^3(j,k;
\al,\be,\ga)[\hat{S}_{\gamma}^{j}, \hat{S}_{0}^{k}]_{+} \right)\,,
$$

$$c_{1}^1(j,j; \al,\be,\ga)=
\frac{\varphi_{\gamma}^{\hbar}(x_{j})\varphi_{0}^{\hbar}(x_{j})}
{\varphi_{\alpha}^{\hbar}(x_{j})\varphi_{\beta}^{\hbar}(x_{j})}\,,
\ \ \
 c_{1}^2(j,k; \al,\be,\ga)=
\frac{\varphi_{\gamma}^{\hbar}(x_{k}) \varphi_{0}^{\hbar}(x_{j})}
{\varphi_{\alpha}^{\hbar}(x_{j})\varphi_{\beta}^{\hbar}(x_{j})}\,,
\ \ \ c_{1}^3(j,k; \al,\be,\ga)=
\frac{\varphi_{\gamma}^{\hbar}(x_{j})\varphi_{0}^{\hbar}(x_{k})}
{\varphi_{\alpha}^{\hbar}(x_{j})\varphi_{\beta}^{\hbar}(x_{j})},$$\\

\beq{qc27}
 [\hat{S}_{\alpha}^{j}, \hat{S}_{0}^{j}]_{-} = i \varepsilon_{\alpha \beta
\gamma} \frac{J_{\beta} - J_{\gamma}}{J_{\alpha}} c_{2}^1(j,j;
\al,\be,\ga) [\hat{S}_{\beta}^{j}, \hat{S}_{\gamma}^{j}]_{+} +
 \eq
$$
 + \sum_{k \neq j}  i \varepsilon_{\alpha \beta \gamma}
\frac{1}{k_{\alpha}J_{\alpha}}
 \cdot \left( c_{2}^2(j,k; \al,\be,\ga) D(\alpha,\beta)
[\hat{S}_{\gamma}^{j}, \hat{S}_{\beta}^{k}]_{+} - c_{2}^3(j,k;
\al,\be,\ga) D(\alpha,\gamma)[\hat{S}_{\gamma}^{k},
\hat{S}_{\beta}^{j}]_{+}
 \right),
$$

$$c_{2}^1(j,j; \al,\be,\ga) =
\frac{\varphi_{\gamma}^{\hbar}(x_{j})\varphi_{\beta}^{\hbar}(x_{j})}
{\varphi_{\alpha}^{\hbar}(x_{j})\varphi_{0}^{\hbar}(x_{j})}\,, \ \
\ c_{2}^2(j,k; \al,\be,\ga)=
 \frac{\varphi_{\gamma}^{\hbar}(x_{j})\varphi_{\beta}^{\hbar}(x_{k})}
{\varphi_{\alpha}^{\hbar}(x_{j})\varphi_{0}^{\hbar}(x_{j})}\,, \ \
\ c_{2}^3(j,k; \al,\be,\ga)=
\frac{\varphi_{\gamma}^{\hbar}(x_{k})\varphi_{\beta}^{\hbar}(x_{j})}
{\varphi_{\alpha}^{\hbar}(x_{j}) \varphi_{0}^{\hbar}(x_{j})}\,,
$$

$$
D(\alpha,\beta)=
 \left(
k_{\alpha}(k_{\alpha}-k_{\beta}-ln^{'}\varphi_{\alpha}^{\hbar}(x_{jk}))
+ln^{'}\frac{\varphi_{\alpha}^{\hbar}(x_{jk})}
{\varphi_{0}^{\hbar}(x_{jk})}(\partial_{z} - E_{1}(\hbar)) \right)
\varphi_{\beta}^{\hbar}(x_{jk}),$$

$$
ln^{'}\varphi_{u}^{\hbar}(x) =
\frac{\partial_{u}(\varphi_{u}^{\hbar}(x))}{\varphi_{u}^{\hbar}(x)},
 $$

 \textit{and for $k \neq j$}:

\beq{qc28}
\begin{array}{c} [\hat{S}_{\alpha}^{j}, \hat{S}_{\beta}^{k}]_{-} =
i \varepsilon_{\alpha \beta \gamma} \frac{1}{k_{\gamma}} \left(
\varphi_{\beta}(x_{jk}) c_{3}^1(j,k; \al,\be,\ga)
[\hat{S}_{\gamma}^{j}, \hat{S}_{0}^{k}]_{+} -
\varphi_{\alpha}(x_{jk}) c_{3}^2(j,k;
\al,\be,\ga)[\hat{S}_{\gamma}^{k}, \hat{S}_{0}^{j}]_{+} \right),
\end{array}
 \eq

$$c_{3}^1(j,k; \al,\be,\ga)=
\frac{\varphi_{\gamma}^{\hbar}(x_{j})\varphi_{0}^{\hbar}(x_{k})}
{\varphi_{\alpha}^{\hbar}(x_{j})\varphi_{\beta}^{\hbar}(x_{k})}\,,
\ \ \ c_{3}^2(j,k; \al,\be,\ga)=
\frac{\varphi_{\gamma}^{\hbar}(x_{k})\varphi_{0}^{\hbar}(x_{j})}
{\varphi_{\alpha}^{\hbar}(x_{j})\varphi_{\beta}^{\hbar}(x_{k})},$$\\

\beq{qc29}
\begin{array}{c} [\hat{S}_{\alpha}^{j}, \hat{S}_{0}^{k}]_{-} =
i \varepsilon_{\alpha \beta \gamma} \frac{1}{k_{\alpha}} \left(
\varphi_{\beta}(x_{jk}) c_{4}^1(j,k; \al,\be,\ga)
[\hat{S}_{\gamma}^{j}, \hat{S}_{\beta}^{k}]_{+} -
\varphi_{\alpha}(x_{jk}) c_{4}^2(j,k;
\al,\be,\ga)[\hat{S}_{\gamma}^{k}, \hat{S}_{\beta}^{j}]_{+}
\right),
\end{array}
 \eq

$$c_{4}^1(j,k; \al,\be,\ga)=\frac{\varphi_{\gamma}^{\hbar}(x_{j})\varphi_{\beta}^{\hbar}(x_{k})}
{\varphi_{\alpha}^{\hbar}(x_{j})\varphi_{0}^{\hbar}(x_{k})}\,, \ \
\ c_{4}^2(j,k;
\al,\be,\ga)=\frac{\varphi_{\gamma}^{\hbar}(x_{k})\varphi_{\beta}^{\hbar}(x_{j})}
{\varphi_{\alpha}^{\hbar}(x_{j})\varphi_{0}^{\hbar}(x_{k})},$$\\

\textit{where}
$$k_{\gamma} = E_{1}(\bfga + \hbar) - E_{1}(\bfga) -
E_{1}(\hbar),$$
$$ J_{\gamma}=
E_{2}(\bfga + \hbar)-E_{2}(\hbar)$$ (see Appendix B),
$$x_{jk} = x_{j} - x_{k}.$$

\textit{Proof.}

Put (\ref{qc5}) in (\ref{qc4}) in the case $N=2$, check the
balance in front of two type fixed matrix elements
$\sigma_{\alpha} \otimes \sigma_{\beta}$ and $\sigma_{\alpha}
\otimes \sigma_{0}$ in left hand side (lhs) and right hand side
(rhs). We fix these elements and compare the coefficients at the
corresponding poles. We get the following expressions for
brackets:

$$
[\hat{S}_{\alpha}^{j}, \hat{S}_{\beta}^{j}]_{-} = i
\varepsilon_{\alpha \beta \gamma} \cdot
\frac{\varphi_{\gamma}^{\hbar}(x_{j})\varphi^{\hbar}(x_{j})}
{\varphi_{\alpha}^{\hbar}(x_{j})\varphi_{\beta}^{\hbar}(x_{j})}
 [\hat{S}_{\gamma}^{j},
\hat{S}_{0}^{j}]_{+} +$$ \beq{qc30}
 + \sum_{k \neq j}
\frac{\varphi_{\gamma}^{\hbar}(x_{jk})}{f^{\hbar}(\alpha,\beta,0)}
\cdot
\frac{\varphi_{\alpha}^{\hbar}(x_{j})\varphi_{\beta}^{\hbar}(x_{k})}
{\varphi_{\alpha}^{\hbar}(x_{j})\varphi_{\beta}^{\hbar}(x_{j})}
[\hat{S}_{\alpha}^{j},\hat{S}_{\beta}^{k}]_{-} -
\frac{\varphi_{\alpha}^{\hbar}(x_{jk})}{f^{\hbar}(\alpha,\beta,0)}
\cdot
\frac{\varphi_{\alpha}^{\hbar}(x_{k})\varphi_{\beta}^{\hbar}(x_{j})}
{\varphi_{\alpha}^{\hbar}(x_{j})\varphi_{\beta}^{\hbar}(x_{j})}
[\hat{S}_{\alpha}^{k}, \hat{S}_{\beta}^{j}]_{-} +
 \eq
$$
+ i \varepsilon_{\alpha \beta \gamma}
\frac{\varphi_{0}^{\hbar}(x_{jk})}{f^{\hbar}(\alpha,\beta,0)}
\cdot
\frac{\varphi_{\gamma}^{\hbar}(x_{k})\varphi_{0}^{\hbar}(x_{j})}
{\varphi_{\alpha}^{\hbar}(x_{j})\varphi_{\beta}^{\hbar}(x_{j})}
[\hat{S}_{\gamma}^{j},\hat{S}_{0}^{k}]_{+} - i \varepsilon_{\alpha
\beta \gamma}
\frac{\varphi_{\gamma}^{\hbar}(x_{jk})}{f^{\hbar}(\alpha,\beta,0)}
\cdot
\frac{\varphi_{\gamma}^{\hbar}(x_{j})\varphi_{0}^{\hbar}(x_{k})}
{\varphi_{\alpha}^{\hbar}(x_{j})\varphi_{\beta}^{\hbar}(x_{j})}
[\hat{S}_{\gamma}^{k}, \hat{S}_{0}^{j}]_{+}\,,
$$

$$
 [\hat{S}_{\alpha}^{j}, \hat{S}_{0}^{j}]_{-} = i \varepsilon_{\alpha \beta
\gamma} \frac{J_{\beta} - J_{\gamma}}{J_{\alpha}} \cdot
\frac{\varphi_{\gamma}^{\hbar}(x_{j})\varphi_{\beta}^{\hbar}(x_{j})}
{\varphi_{\alpha}^{\hbar}(x_{j})\varphi_{0}^{\hbar}(x_{j})}
 [\hat{S}_{\beta}^{j},
\hat{S}_{\gamma}^{j}]_{+} +$$ \beq{qc31}
 + \sum_{k \neq j}
\frac{\varphi_{0}^{\hbar}(x_{jk})}{J_{\alpha}} \cdot
\frac{\varphi_{0}^{\hbar}(x_{k})} {\varphi_{0}^{\hbar}(x_{j})}
[\hat{S}_{\alpha}^{j},\hat{S}_{0}^{k}]_{-} +
\frac{\varphi_{\alpha}^{\hbar}(x_{jk})}{J_{\alpha}} \cdot
\frac{\varphi_{\alpha}^{\hbar}(x_{k})}
{\varphi_{\alpha}^{\hbar}(x_{j})}
[\hat{S}_{\alpha}^{k},\hat{S}_{0}^{j}]_{-} + \eq
$$
- i \varepsilon_{\alpha \beta \gamma}
\frac{\varphi_{\gamma}^{\hbar}(x_{jk})}{J_{\alpha}} \cdot
\frac{\varphi_{\gamma}^{\hbar}(x_{k})\varphi_{\beta}^{\hbar}(x_{j})}
{\varphi_{\alpha}^{\hbar}(x_{j})\varphi_{0}^{\hbar}(x_{j})}
[\hat{S}_{\beta}^{j},\hat{S}_{\gamma}^{k}]_{+} + i
\varepsilon_{\alpha \beta \gamma}
\frac{\varphi_{\beta}^{\hbar}(x_{jk})}{J_{\alpha}} \cdot
\frac{\varphi_{\gamma}^{\hbar}(x_{j})\varphi_{\beta}^{\hbar}(x_{k})}
{\varphi_{\alpha}^{\hbar}(x_{j})\varphi_{0}^{\hbar}(x_{j})}
 [\hat{S}_{\gamma}^{j}, \hat{S}_{\beta}^{k}]_{+}\,,
$$

 \textit{and for $k \neq j$}:

\beq{qc32}
 [\hat{S}_{\alpha}^{j}, \hat{S}_{\beta}^{k}]_{-} =
\frac{\varphi_{\gamma}^{\hbar}(x_{jk})}{\varphi_{0}^{\hbar}(x_{jk})}
\cdot
\frac{\varphi_{\alpha}^{\hbar}(x_{k})\varphi_{\beta}^{\hbar}(x_{j})}
{\varphi_{\alpha}^{\hbar}(x_{j})\varphi_{\beta}^{\hbar}(x_{k})}
[\hat{S}_{\alpha}^{k}, \hat{S}_{\beta}^{j}]_{-} + \eq
$$
+ i \varepsilon_{\alpha \beta \gamma}
\frac{\varphi_{\alpha}^{\hbar}(x_{jk})}{\varphi_{0}^{\hbar}(x_{jk})}
\cdot
\frac{\varphi_{\gamma}^{\hbar}(x_{k})\varphi_{0}^{\hbar}(x_{j})}
{\varphi_{\alpha}^{\hbar}(x_{j})\varphi_{\beta}^{\hbar}(x_{k})}
[\hat{S}_{\gamma}^{k},\hat{S}_{0}^{j}]_{+} - i \varepsilon_{\alpha
\beta \gamma}
\frac{\varphi_{\beta}^{\hbar}(x_{jk})}{\varphi_{0}^{\hbar}(x_{jk})}
\cdot
\frac{\varphi_{\gamma}^{\hbar}(x_{j})\varphi_{0}^{\hbar}(x_{k})}
{\varphi_{\alpha}^{\hbar}(x_{j})\varphi_{\beta}^{\hbar}(x_{k})}
[\hat{S}_{\gamma}^{j}, \hat{S}_{0}^{k}]_{+}\,,
\\
$$

\beq{qc33}
 [\hat{S}_{\alpha}^{j}, \hat{S}_{0}^{k}]_{-} =
\frac{\varphi_{\alpha}^{\hbar}(x_{jk})}{\varphi_{0}^{\hbar}(x_{jk})}
\cdot
\frac{\varphi_{\alpha}^{\hbar}(x_{k})\varphi_{0}^{\hbar}(x_{j})}
{\varphi_{\alpha}^{\hbar}(x_{j})\varphi_{0}^{\hbar}(x_{k})}
[\hat{S}_{\alpha}^{k}, \hat{S}_{0}^{j}]_{-} + \eq
$$
+ i\varepsilon_{\alpha \beta \gamma}
\frac{\varphi_{\gamma}^{\hbar}(x_{jk})}{\varphi_{0}^{\hbar}(x_{jk})}
\cdot
\frac{\varphi_{\beta}^{\hbar}(x_{j})\varphi_{\gamma}^{\hbar}(x_{k})}
{\varphi_{\alpha}^{\hbar}(x_{j})\varphi_{0}^{\hbar}(x_{k})}
[\hat{S}_{\beta}^{j},\hat{S}_{\gamma}^{k}]_{+} - i
\varepsilon_{\alpha \beta \gamma}
\frac{\varphi_{\beta}^{\hbar}(x_{jk})}{\varphi_{0}^{\hbar}(x_{jk})}
\cdot
\frac{\varphi_{\beta}^{\hbar}(x_{k})\varphi_{\gamma}^{\hbar}(x_{j})}
{\varphi_{\alpha}^{\hbar}(x_{j})\varphi_{0}^{\hbar}(x_{k})}
[\hat{S}_{\beta}^{k}, \hat{S}_{\gamma}^{j}]_{+}\,.
$$

It is possible to express all commutators by the anti-commutators.
In fact for the brackets $[\hat{S}_{\alpha}^{j},
\hat{S}_{\beta}^{k}]_{-}, \ [\hat{S}_{\alpha}^{j},
\hat{S}_{0}^{k}]_{-}$ we have two additional equations
(permutation $j \leftrightarrow k$). Solving the system of six
equations we get (\ref{qc26})-(\ref{qc29}). $\Box$

\subsection{Quantum Determinant}
In this section we prove for ${\rm GL}(2,\mathbb{C})$ that the
quantum determinant generates central elements of the exchange
algebra \beq{q1}
R_{12}(z_1,z_2)\hat{L}_1(z_1)\hat{L}_2(z_2)=\hat{L}_2(z_2)\hat{L}_1(z_1)R_{12}(z_1,z_2)\eq
for R and L defined in (\ref{qc22}) and (\ref{qc5}).

Let us start from the classical algebra (\ref{gen})-(\ref{3.7}).
To prove in ${\rm GL}(N,\mathbb{C})$ case that $\det L(z)$
generates the Casimir functions of the Poisson structure
(\ref{gen})-(\ref{3.7}) consider each side of the equality
$$
\{L_1(z)\dots L_N(z),L_{N+1}(w)\}=[L_1(z)\dots
L_N(z)L_{N+1}(w),r_{1,N+1}(z,w)+\dots+r_{N,N+1}(z,w)]
$$
as a linear operator acting on $\bigotimes\limits_{i=1}^{N+1}V_i$,
where $V_i\cong\mathbb{C}^N$ are vector spaces, $L_i\in \hbox{End}
V_i$ and $r_{ik}\in \hbox{End} (V_i\otimes V_k)$. The determinant
$\det L(z)$ obviously appears on the subspace
$\left[\bigwedge\limits_{i=1}^N V_i\right]\otimes V_{N+1}$. The
r.h.s. on this subspace reduces to the following:
$$
[\det L(z) \cdot L_{N+1}(w),
Tr_1r_{1,N+1}(z,w)+Tr_Nr_{N,N+1}(z,w)].
$$
Here traces $Tr_i$ are taken over $\hbox{End} V_i$ components. All
of them vanish for the r-matrix (\ref{rmat}). End of the proof for
the classical case.

In quantum case the determinant is replaced by the quantum
determinant:
$$
\det_\hbar=\tr(P^-\hat{L}(z,\hbar)\otimes\hat{L}(z+2\hbar,\hbar))\,,
$$
where $P^-$ is the projection into skewsymmetric part of the
tensor product:
$$
P^-a\otimes b=\frac{1}{2}\left(a\otimes b-b\otimes a\right)\,.
$$

Here we discuss only $2\times 2$ case. The R-matrix
$$
R_{12}(z,w)=\sum\limits_{a=0}^3\varphi_a^\eta(z-w)\sigma_a\otimes\sigma_a
$$
satisfies the following important condition:
$$
R_{12}(z,z+2\hbar)=4\frac{\vth'(0)}{\vth(2\hbar)}P^-,
$$

and $$\ P^-=\frac{1}{4}\left(1\otimes
1-\sum\limits_{\al=1}^3\sigma_\al\otimes \sigma_\al\right)\,.
$$

Consider the product $L_1(z_1)L_2(z_2)L_3(w)\in V^{\otimes 3}$.

 It
follows from the Yang-Baxter equation that
$$
R_{12}R_{13}R_{23}\hat{L}_1\hat{L}_2\hat{L}_3=\hat{L}_3\hat{L}_2\hat{L}_1R_{12}R_{13}R_{23}
$$
Put $z_2=z_1+2\hbar$. Then
$$
P^-_{12}R_{13}R_{23}\hat{L}_1\hat{L}_2\hat{L}_3=\hat{L}_3\hat{L}_2\hat{L}_1P^-_{12}R_{13}R_{23}
$$
The next statement is the most important one:
$$
P^-_{12}R_{13}R_{23}\sim P^-_{12}\otimes 1_3
$$
It follows from direct calculations. For the simplicity one can
use the following identity for $\al,\be,\ga\sim 1,2,3$ up to the
cyclic permutations: \beq{idt}
-\varphi_0^\hbar(x)\varphi_\gamma(x-2\hbar)+
\varphi_\ga^\hbar(x)\varphi_0(x-2\hbar)+
\varphi_\be^\hbar(x)\varphi_\al(x-2\hbar)+
\varphi_\al^\hbar(x)\varphi_\be(x-2\hbar)=0 \eq Using also a
simple fact $Tr_{12}\left(P_{12}^-\hat{L}_1\hat{L}_2\right)=
Tr_{12}\left(P_{12}^-\hat{L}_2\hat{L}_1\right)$ we come to the
final result:
$$
[Tr_{12}\left(P_{12}^-\hat{L}_1(z-2\hbar)\hat{L}_2(z)\right),\hat{L}_3(w)]=0
$$

\subsection{Nonhomogeneous algebra and Reflection Equation}

Consider the rank two case $(N=2)$ with four marked points $n=4$.
As an initial data we put the marked points on $z=0$ and the
half-periods of $\Si_\tau$
$$
x_0=0\,,~x_1=\frac{\tau}{2}=\om_2\,,~x_2=\frac{1+\tau}{2}=\om_1+\om_2\,,~
x_3=\frac{1}{2}=\om_1\,,
$$
and assume that \beq{id}
S^j_\al=\de^j_\al\ti{\nu}_\al\,,~(j=1\,,2\,,3)\,, \eq while
$S^0_\al=S_\al$ are arbitrary. This choice appears as a
consequence of the reduction $L(z)L(-z)=1\times \det L(z)$.

Let  $R^-$ be the quantum vertex R-matrix, that arises in the XYZ
model. We introduce also the matrix $R^+$ \beq{60} R^\pm(z,w)=
\sum\limits_{a=0}^3\varphi^\frac{\hbar}{2}_a(z\pm w) \sigma_a
\otimes \sigma_a \,.\eq

Define the quantum Lax operator \beq{61}
\hat{L}(z)=\hat{S}_0\phi^\hbar(z)\si_0+
\sum_\al(\hat{S}_\al\vf^\hbar_\al(z)+
\tilde{\nu}_\al\vf^\hbar_\al(z-\om_\al))\si_\al\,. \eq
\begin{predl}
The Lax operator satisfies the quantum reflection equation
\beq{62} R^{-}(z,w)\hat{L}_1(z)R^{+}(z,w)\hat{L}_2(w)=
\hat{L}_2(w)R^{+}(z,w)\hat{L}_1(z)R^{-}(z,w)\,, \eq if its
components $S_a$ generate the associative algebra with relations:
\beq{63}
[\tilde{\nu}_\al,\tilde{\nu}_\be]=0\,,~~[\tilde{\nu}_\al,\hat{S}_a]=0\,,
\eq \beq{64}
i[\hat{S}_0,\hat{S}_\al]_+=[\hat{S}_\be,\hat{S}_\ga]\,, \eq
\beq{65}
[\hat{S}_\ga,\hat{S}_0]=i\frac{K_\be-K_\al}{K_\ga}[\hat{S}_\al,\hat{S}_\be]_+
-2i\f1{K_\ga}(\tilde{\nu}_\al\rho_\al\hat{S}_\be-
\tilde{\nu}_\be\rho_\be\hat{S}_\al)\,, \eq where
$$
K_\al=E_1(\hbar+\check{\al})-E_1(\hbar)-E_1(\check{\al})\,,~~
\rho_\al=-\exp(-2\pi
\imath\check{\al}\p_\tau\check{\al})\phi(\check{\al}+\hbar,-\check{\al})\,.
$$
\end{predl}
The proof is based on the direct check. Details can be found in
\cite{LOZ}.

If all $\nu_\al=0$ (\ref{63}) -- (\ref{65}) the algebra coincides
with the Sklyanin algebra. Therefore, the algebra (\ref{63}) --
(\ref{65}) is a three parametric deformation of the Sklyanin
algebra.

\bigskip
Two elements
$$
C_1=\hat{S}_0^2+\sum_{\al}\hat{S}_\al^2\,,
$$
$$
C_2=\sum_{\al}
\hat{S}_\al^2K_\al(K_\al-K_\be-K_\ga)+2\ti{\nu}_\al\rho_\al
K_\al\hat{S}_\al
$$
belong to the center of the generalized Sklyanin algebra
(\ref{63}), (\ref{64}). They are the coefficients of the expansion
of the quantum determinant
$$
\det_\hbar=\tr(P^-\hat{L}(z,\hbar)\otimes\hat{L}(z+2\hbar,\hbar))\,.
$$


\section{Appendix}
\subsection{Appendix A. Elliptic functions.}
\setcounter{equation}{0}
\def\theequation{A.\arabic{equation}}

We assume that $q=\exp 2\pi i\tau$, where $\tau$ is the modular
parameter of the elliptic curve $E_\tau$.

The basic element is the theta  function: \beq{A.1a}
\vth(z|\tau)=q^{\frac {1}{8}}\sum_{n\in {\bf Z}}(-1)^n\bfe(\oh
n(n+1)\tau+nz)=~~ (\bfe=\exp 2\pi\imath) \eq

\bigskip

{\it The  Eisenstein functions} \beq{A.1}
E_1(z|\tau)=\p_z\log\vth(z|\tau),
~~E_1(z|\tau)\sim\f1{z}-2\eta_1z, \eq where \beq{A.6}
\eta_1(\tau)=\frac{24}{2\pi i}\frac{\eta'(\tau)}{\eta(\tau)}\,,~~
\eta(\tau)=q^{\frac{1}{24}}\prod_{n>0}(1-q^n)\,. \eq is the
Dedekind function. \beq{A.2} E_2(z|\tau)=-\p_zE_1(z|\tau)=
\p_z^2\log\vth(z|\tau), ~~E_2(z|\tau)\sim\f1{z^2}+2\eta_1\,. \eq

{\it Relation to the Weierstrass functions} \beq{a100}
\zeta(z,\tau)=E_1(z,\tau)+2\eta_1(\tau)z\,,
~~\wp(z,\tau)=E_2(z,\tau)-2\eta_1(\tau)\,. \eq The highest
Eisenstein functions \beq{A.2a}
E_j(z)=\frac{(-1)^j}{(j-1)!}\p^{(j-2)}E_2(z)\,,~~(j>2)\,. \eq

The next important function is \beq{A.3} \phi(u,z)= \frac
{\vth(u+z)\vth'(0)} {\vth(u)\vth(z)}\,. \eq \beq{A.300}
\phi(u,z)=\phi(z,u)\,,~~\phi(-u,-z)=-\phi(u,z)\,. \eq It has a
pole at $z=0$ and \beq{A.3a}
\phi(u,z)=\frac{1}{z}+E_1(u)+\frac{z}{2}(E_1^2(u)-\wp(u))+\ldots\,.
\eq

\beq{A3c} \p_u\phi(u,z)=\phi(u,z) (E_1(u+z)-E_1(u)) \,. \eq

\beq{A3e} \p_z\phi(u,z)=\phi(u,z) (E_1(u+z)-E_1(z)) \,. \eq

\beq{A3d} \lim_{z\to 0}\ln\p_u\phi(u,z)=-E_2(u) \,. \eq

{\it Heat equation} \beq{A.4b} \p_\tau\phi(u,w)-\f1{2\pi
i}\p_u\p_w\phi(u,w)=0\,. \eq

{\it Quasi-periodicity}

\beq{A.11} \vth(z+1)=-\vth(z)\,,~~~\vth(z+\tau)=-q^{-\oh}e^{-2\pi
iz}\vth(z)\,, \eq \beq{A.12}
E_1(z+1)=E_1(z)\,,~~~E_1(z+\tau)=E_1(z)-2\pi i\,, \eq \beq{A.13}
E_2(z+1)=E_2(z)\,,~~~E_2(z+\tau)=E_2(z)\,, \eq \beq{A.14}
\phi(u,z+1)=\phi(u,z)\,,~~~\phi(u,z+\tau)=e^{-2\pi \imath
u}\phi(u,z)\,. \eq \beq{A.15}
\p_u\phi(u,z+1)=\p_u\phi(u,z)\,,~~~\p_u\phi(u,z+\tau)=e^{-2\pi
\imath u}\p_u\phi(u,z)-2\pi\imath\phi(u,z)\,. \eq

 {\it  The Fay three-section formula:}
\beq{ad3}
\phi(u_1,z_1)\phi(u_2,z_2)-\phi(u_1+u_2,z_1)\phi(u_2,z_2-z_1)-
\phi(u_1+u_2,z_2)\phi(u_1,z_1-z_2)=0\,. \eq

{\it  From (\ref{A3e}) and (\ref{ad3}) we have:}

\beq{ad31} \phi(u_1,z)\phi(u_2,z)=\phi(u_1+u_2,z)(E_1(u_1)+
E_1(u_2) - E_1(u_1 + u_2 + z) + E_1(z)) \,. \eq

Particular cases of this formula are the  functional equations
\beq{ad2}
\phi(u,z)\p_v\phi(v,z)-\phi(v,z)\p_u\phi(u,z)=(E_2(v)-E_2(u))\phi(u+v,z)\,,
\eq \beq{i}
\phi(u,z_1)\phi(-u,z_2)=\phi(u,z_1-z_2)(-E_1(z_1)+E_1(z_2)-E_{1}(u)+E_{1}(u+z_{1}-z_{2}))=
\eq
$$
 = \phi(u,z_1-z_2)(-E_1(z_1)+E_1(z_2)+\p_u\phi(u,z_2-z_1))\,,
$$
\beq{ir1} \phi(u,z)\phi(-u,z)=E_2(z)-E_2(u)\,.
 \eq

\beq{ir} \phi(v,z-w)\phi(u_1-v,z)\phi(u_2+v,w)
-\phi(u_1-u_2-v,z-w)\phi(u_2+v,z)\phi(u_1-v,w)= \eq
$$
\phi(u_1,z)\phi(u_2,w)f(u_1,u_2,v)\,,
$$
where \beq{ir8} {\bf
f}(u_1,u_2,v)=E_1(v)-E_1(u_1-u_2-v)+E_1(u_1-v)-E_1(u_2+v)\,. \eq
One can rewrite the last function as \beq{ir3} {\bf
f}(u_1,u_2,v)=-\frac{ \vth'(0)\vth(u_1)\vth(u_2)\vth(u_2-u_1+2v)
}{ \vth(u_1-v)\vth(u_2+v)\vth(u_2-u_1+v)\vth(v) }\,. \eq

Using (\ref{A.1}), (\ref{A.2}), (\ref{A.3a}) one can derive from
(\ref{ir}) some important particular cases. One of them
corresponding to $v=u_1$ (or $v=-u_2$), is the Fay identity
(\ref{ad3}). Another particular case comes from $u_1=0$ (or
$u_2=u$): \beq{ir7}
\phi(v,z-w)\phi(-v,z)\phi(u+v,w)-\phi(-u-v,z-w)\phi(u+v,z)\phi(-v,w)=
\eq
$$
\phi(u_1,z)(E_2(u+v)-E_2(v))\,.
$$
If $u_2\to -v$ then (\ref{ir}) in the first non-trivial order take
the form for $u_1=\al,~u_2=\be$ \beq{ir4}
\phi(-\be,z-w)E_1(w)\phi(\al+\be,z)-
\phi(\al,z-w)E_1(z)\phi(\al+\be,w)= \eq
$$
\phi(\al,z)\phi(\be,w)(E_1(\al)+E_1(\be)-E_1(\al+\be))\,.
$$

\subsection{Appendix B.  Lie algebra $\sln$ and elliptic functions}
\setcounter{equation}{0}
\def\theequation{B.\arabic{equation}}

Introduce the notation
$$
{\bf e}_N(z)=\exp (\frac{2\pi i}{N} z)
$$
 and two matrices
\beq{q} Q=\di({\bf e}_N(1),\ldots,{\bf e}_N(m),\ldots,1) \eq
\beq{la} \La=\de_{j,j+1}\,,~~(j=1,\ldots,N\,,~mod\,N)\,.
 \eq
 Let
\beq{B.10}
\mZ^{(2)}_N=(\mZ/N\mZ\oplus\mZ/N\mZ)\,,~~\ti{\mZ}^{(2)}_N)=
\mZ^{(2)}_N\setminus(0,0) \eq be the two-dimensional lattice of
order $N^2$ and $N^2-1$ correspondingly. The matrices
$Q^{a_1}\La^{a_2}$, $a=(a_1,a_2)\in\mZ^{(2)}_N$ generate a basis
in the group $\GLN$, while $Q^{\al_1}\La^{\al_2}$,
$\al=(\al_1,\al_2)\in\ti{\mZ}^{(2)}_N$ generate a basis in the Lie
algebra $\sln$. More exactly, we introduce the following basis in
$\GLN$. Consider the projective representation of $\mZ^{(2)}_N$ in
$\GLN$ \beq{B.11} a\to T_{a}= \frac{N}{2\pi
i}\bfe_N(\frac{a_1a_2}{2})Q^{a_1}\La^{a_2}\,, \eq \beq{AA3a}
T_aT_b=\frac{N}{2\pi i}\bfe_N(-\frac{a\times b}{2})T_{a+b}\,, ~~
(a\times b=a_1b_2-a_2b_1) \,. \eq Here $\frac{N}{2\pi i}
\bfe_N(-\frac{a\times b}{2})$ is a non-trivial two-cocycle in
$H^2(\mZ^{(2)}_N,\mZ_{2N})$. The matrices $T_\al$,
$\al\in\ti{\mZ}^{(2)}_N$ generate a basis in $\sln$. It follows
from (\ref{AA3a}) that \beq{AA3b}
[T_{\al},T_{\be}]=\bfC(\al,\be)T_{\al+\be}\,, \eq where
$\bfC(\al,\be)=\frac{N}{\pi}\sin\frac{\pi}{N}(\al\times \be)$ are
 the structure constants of $\sln$.

 For $N=2$ the basis $T_{\al}$ is proportional to the basis of the Pauli
matrices:
 $$
 T_{(1,0)}=\f1{\pi\imath}\si_3\,,~~
 T_{(0,1)}=\f1{\pi\imath}\si_1\,,~~
  T_{(1,1)}=\f1{\pi\imath}\si_2\,.
  $$

 The Lie coalgebra $\gg^*=\sln$ has the dual basis
  \beq{db}
 \gg^*=\{\bfS=\sum_{\ti{\mZ}^{(2)}_N}S_\ga t^\ga\}\,,~~
 t^\ga=\frac{2\pi\imath}{N^2}T_{-\ga}\,,~~\lan T_\al
t^\be\ran=\de_{\al}^{-\be}\,.
 \eq
It follows from (\ref{AA3b}) that $\gg^*$ is a Poisson space with
the linear brackets \beq{A101}
\{S_\al,S_\be\}=\bfC(\al,\be)S_{\al+\be}\,. \eq The coadjoint
action in these basises   takes the form \beq{coad} {\rm
ad}^*_{T_\al}t^\be=\bfC(\al,\be)t^{\al+\be}\,. \eq

Let $\bfga=\frac{\ga_1+\ga_2\tau}{N}$. Then introduce the
following  constants on $\ti{\mZ}^{(2)}$: \beq{AA50}
\vth(\bfga)=\vth\bigl(\frac{\ga_1+\ga_2\tau}{N}\bigr)\,, ~~
E_1(\bfga)=E_1\bigl(\frac{\ga_1+\ga_2\tau}{N}\bigr)\,,
~~E_2(\bfga)=E_2\bigl(\frac{\ga_1+\ga_2\tau}{N}\bigr)\,, \eq
\beq{ph} \phi_\ga(z)=\phi(\bfga,z)\,, \eq \beq{vf}
\vf_\ga(z)=\bfe_N(\ga_2z)\phi_\ga(z)\,, \eq \beq{vf1}
\vf_{\gamma,\eta}(z)=\bfe_N(\ga_2z)\phi(\eta +
\frac{\ga_1+\ga_2\tau}{N},z)\,. \eq They have the following
quasi-periodicities \beq{qpe1}
\vf_\ga(z+1)=\bfe_N(\ga_2)\vf_\ga(z)\,,~~
\vf_\ga(z+\tau)=\bfe_N(-\ga_1)\vf_\ga(z)\,, \eq \beq{qpe2}
\vf_{\ga,\eta}(z+1)=\bfe_N(\ga_2)\vf_{\ga,\eta}(z)\,,~~
\vf_{\ga,\eta}(z+\tau)=\bfe_N(-\ga_1-\eta)\vf_{\ga,\eta}(z)\,, \eq

The important relations for these functions are \beq{ir21}
\frac{\varphi_{\eta}(z_{1}-z_{2})}{\varphi_{\eta}(z_{1})} =
\frac{1}{\varphi_{\eta}(z_{2})} (E_{1}(z_{2})+ E_{1}(\eta) +
E_{1}(z_{1}-z_{2}) - E_{1}(z_{1}+\eta) ),
 \eq
\beq{ir22}
\frac{\varphi_{\alpha,\eta}(z_{1}-z_{2})}{\varphi_{\eta}(z_{1})} =
\frac{1}{\varphi_{\eta}(z_{2})} \varphi_{\alpha}(z_{1}-z_{2}) +
\frac{\varphi_{-\alpha}(z_{2})}{\varphi_{\eta}(z_{2})\varphi_{\alpha}(\eta)}
\varphi_{\alpha}(z_{1}+\eta),
 \eq
\beq{ir23} \frac{\varphi_{\alpha,\eta}(z)}{\varphi_{\eta}(z)} =
\frac{\varphi_{\alpha}(z+\eta)}{\varphi_{\alpha}(\eta)}.
 \eq

Another important relation in the case $N=2$ is \beq{ir24}
 k_{\gamma} f^{\hbar}(\gamma,\alpha,0) = J_{\gamma}=
E_{2}(\gamma + \hbar)-E_{2}(\hbar),
 \eq
where
$$
k_{\gamma} = E_{1}(\gamma + \hbar) - E_{1}(\gamma) - E_{1}(\hbar),
$$
We give a short comment of this formula. From
(\ref{ir}),(\ref{ir8}) we have: \beq{ir25}
 (E_{1}(\gamma + \hbar) - E_{1}(\gamma) - E_{1}(\hbar))
(E_{1}(\alpha + \hbar) + E_{1}(-\beta + \hbar) - E_{1}(\gamma +
\hbar) - E_{1}(\hbar)) = E_{2}(\gamma + \hbar)-E_{2}(\hbar), \eq

where we suppose $\alpha-\beta=\gamma$. The function at r.h.s. and
the function at l.h.s. have the coinciding poles ($\hbar=0,
\hbar=-\gamma$) and zeroes ($\hbar = -\frac{1}{2}\gamma$), so we
come to the equality of these functions.

Define the function \beq{f} f_\ga(z)=\bfe_N(\ga_2z)
\p_u\phi(u,z)|_{u=\bfga}=\vf_\ga(z)(E_1(\bfga+z)-E_1(\bfga))\,.
\eq It follows from (\ref{A3c}) that \beq{f1}
f_\ga(z)=\vf_\ga(z)(E_1(\bfga+z)-E_1(\bfga))\,. \eq \beq{fzh}
\bff_{\al,\be,\ga}=E_1(\bfga)-E_1(\bfal-\bfbe-\bfga)+E_1(\bfal-\bfga)
-E_1(\bfbe-\bfga)\,. \eq (see (\ref{ir8}))

It follows from (\ref{A.3}) that \beq{qp}
\vf_\ga(z+1)=\bfe_N(\ga_2)\vf_\ga(z)\,,~~
\vf_\ga(z+\tau)=\bfe_N(-\ga_1)\vf_\ga(z)\,. \eq \beq{qpf}
f_\ga(z+1)=\bfe_N(\ga_2)f_\ga(z)\,,~~
f_\ga(z+\tau)=\bfe_N(-\ga_1)f_\ga(z)-2\pi\imath\vf_\ga(z)\,. \eq
The modification of (\ref{ir}) is \beq{ir5}
\varphi_\ga(z-x_j)\varphi_{-\ga}(z-x_k)=\varphi_\ga(x_k-x_j)
(E_1(z-x_k)-E_1(z-x_j))-f_\ga(x_k-x_j)\,. \eq

\small{

\end{document}